\documentstyle[11pt, epsfig]{article}
\def\be{\begin{equation}}
\def\ee{\end{equation}}
\def\bea{\begin{eqnarray}}
\def\eea{\end{eqnarray}}

\def\e{{\mbox{e}}}
\def\del{\partial}
\def\vr{{\vec r}}
\def\vq{{\vec q}}
\def\vsigma{{\vec \sigma}}
\def\vtau{{\vec \tau}}
\def\hpiNN{{h_{\pi NN}^{(1)}}}
\def\barE{{\tilde {\mbox{E}}}}

\begin{document}

\renewcommand{\thefootnote}{\#\arabic{footnote}}
\setcounter{footnote}{0}
\vskip 0.2cm
\begin{center}
{\Large \bf Asymmetry in $\vec{n} + p \rightarrow d + \gamma$}
\vskip 1.5cm{
{\large Chang Ho Hyun}$^{a,}$\footnote{
e-mail: hch@zoo.snu.ac.kr}\footnote{
Present address: Department of Physics, Sung Kyun Kwan University,
Suwon 440-746, Korea
},
{\large Tae-Sun Park}$^{b,c,}$\footnote{e-mail: tspark@nuc003.psc.sc.edu} and
{\large  Dong-Pil Min}$^{a,}$\footnote{e-mail: dpmin@phya.snu.ac.kr} \\ 
\vskip 0.1cm 
$^a${\it Department of Physics, Seoul National
University, Seoul 151-742, Korea}} \\  
$^b${\it Theory
Group, TRIUMF, 4004 Wesbrook Mall, Vancouver, B.C., Canada V6T 2A3}
\\
$^c${\it Department of Physics and Astronomy,
University of South Carolina, Columbia SC29208, USA}

\end{center}
\vskip 1cm

\centerline{\bf Abstract} \vskip 0.1cm

Heavy-baryon chiral perturbation theory (HBChPT)
is applied to the 
asymmetry $A_\gamma$ 
in $\vec{n} + p \rightarrow d + \gamma$ 
at threshold, which arises
due to the weak parity non-conserving interactions.
Instead of appealing to Siegert's theorem,
transition operators up to 
next-to-leading chiral order are derived
and the corresponding amplitudes are
evaluated with the
Argonne $v_{18}$ wavefunctions.
In addition to the impulse contribution,
both parity-conserving and parity-non-conserving
two-body one-pion-exchange diagrams 
appear up to this order.
Our prediction for the asymmetry is
$A_\gamma = - 0.10\ \hpiNN$,
which is close to the 
Siegert's theorem based result,
$A_\gamma \simeq -0.11\ \hpiNN$.
This illustrates that HBChPT
is effectively applied to the 
parity-non-conserving physics.


\newpage

There are still large discrepancies in the value of the $\hpiNN$,
where $\hpiNN$ is the leading order
weak $\pi NN$ coupling constant.
For example, detection of the circular polarization
of 1081 keV gamma rays from $^{18}$F transition predicts
$|\hpiNN| \leq 1.43 \times 10^{-7}$ \cite{aharnps85}.
Measurement of anapole moment of $^{133}$Cs predicts, on the other hand,
much larger value,
$\hpiNN = (9.5 \pm 0.22 \pm 0.34) \times 10^{-7}$ \cite{fm97}.
The process
\be
\vec{n} + p \rightarrow d + \gamma
\label{process}
\ee
at threshold
is free from the uncertainties of 
many-body ($A \geq 3$) systems,
and thus a suitable source for
the study of the parity non-conserving (PNC)
effects in nuclear reactions.
As a relevant observable for this purpose
the asymmetry $A_\gamma$ is investigated,
which is defined by the dependence of the cross
section on the angle $\theta$ 
defined by the directions of 
the photon emission and the neutron polarization,
$W(\theta)\propto 1 + A_\gamma \cos\theta$.
The first calculation of the $A_\gamma$ was performed by Danilov 
\cite{danilov} obtaining
$A_\gamma = -0.08\ \hpiNN$.
More elaborate attempts were made
with realistic wavefunctions \cite{des75, dmnpa78, ddh80},
\begin{eqnarray}
A_\gamma = -0.11\ \hpiNN.
\label{eq:dm78}
\end{eqnarray}
In more detail, 
Hamada-Johnston, Reid-soft-core 
and Tourreil-Sprung potentials
are adopted to yield
$A_\gamma= -0.109\ \hpiNN$, $-0.114\ \hpiNN$ 
and $-0.107\ \hpiNN$, respectively \cite{des75}.
The available data are from the ILL experiment  
\cite{acanada88}, 
$A^{\rm ILL}_\gamma = -( 1.5 \pm 4.8) \times 10^{-8}$. 
This data with the eq.(\ref{eq:dm78}) then imposes
$\hpiNN = (1.4 \pm 4.4) \times 10^{-7}$.
At LANSCE \cite{snowetal98},
an experiment that aims at having
$10^{-9}$ accuracy in $A_\gamma$ is 
under progress, which will sharpen the
determination of $\hpiNN$ greatly.

The good convergence among the above various calculations
is mostly due to Siegert's theorem \cite{siegert, aspr51}.
It relates the major part of the E1 amplitude at low-energy
to the one-body charge density
whose amplitude can be estimated reliably
without detailed informations of the 
nucleon-nucleon reactions.
This indicates that a substantial departure from
eq.(\ref{eq:dm78}) is unlikely.
In the meantime,
effective field theories (EFTs) have recently gained great successes
in low-energy two-nucleon systems,
which include
the Solar proton fusion \cite{pkmr-astro}, 
the total radiative $np$ capture cross section \cite{pmr-prl95}
and its spin observables \cite{pkmr2000, crs99},
the deuteron properties and low-energy nucleon-nucleon
phase shifts \cite{pkmr, ksw98, egm99, ork96}.
Kaplan, Savage, Springer and Wise
\cite{kssw99} (KSSW)
performed an EFT calculation of the process (\ref{process}),
using the so-called power-divergence subtraction scheme.
They found rather surprising result,
$A_\gamma = 0.17\ \hpiNN$,
where the difference in overall sign is simply due to 
a mismatch in conventions (see Ref. \cite{des065,savage0012}).
Desplanques \cite{des065} has analyzed KSSW's result
in great detail.
He showed that KSSW's result is
-- apart from the overall sign --
exactly equivalent to the
conventional result but with
the zero-range approximation (ZRA) for the wavefunctions.
The ZRA is
responsible for the KSSW's overestimation of the asymmetry.

In this paper, we will show that
the asymmetry can be understood accurately
by HBChPT, an EFT that has been
thoroughly tested in low-energy
nuclear physics.
For this purpose we will go to next-to-leading-order (NLO)
in Weinberg's power counting \cite{wein90},
with the wavefunctions obtained by
the Argonne $v_{18}$ potential \cite{av18}.
So far this {\it hybrid} method has been found
to be quite powerful.
For example,
the total cross section of the neutron thermal capture
was found to agree to the experimental data
perfectly with the theoretical error bar 
about 1\% \cite{pmr-prl95}.

The leading-order PNC Lagrangian 
takes the form \cite{ddh80, ksnpa93, kssw99}
\begin{eqnarray}
{\cal L}_{pnc} =
-\frac{\hpiNN}{\sqrt{2}} \epsilon^{3ab} N^\dagger \tau^a \pi^b N,
\label{eq:pncpinn}
\end{eqnarray}
where $\epsilon^{123}=+1$
and $\hpiNN$ is the weak coupling constant that should be determined
from the experimental value of the asymmetry.
There are other PNC interactions at higher order,
but our current poor understanding of the nuclear PNC effects
does not make much sense to include them.
Thus we will limit ourselves to the above leading order PNC term.
Due to the smallness of $\hpiNN\ (\sim 10^{-7})$,
it is sufficient to consider only the contributions
linear in $\hpiNN$ for the PNC amplitude.
The factor $\hpiNN$ appears either in the wavefunctions 
or in the currents,
\begin{eqnarray}
\delta \langle \Psi_f | \vec{J} |\Psi_i\rangle 
= \langle\delta \Psi_f | \vec{J} |\Psi_i\rangle 
+ \langle\Psi_f | \vec{J} |\delta \Psi_i\rangle 
+ \langle\Psi_f | \delta \vec{J} | \Psi_i\rangle,
\label{dpsi}
\end{eqnarray}
where $\Psi_i$ ($\Psi_f$) and ${\vec J}$ are the
initial (final) wavefunctions and the electromagnetic currents,
respectively.
The ``$\delta$'' marks stand for the
first order perturbation with respect to the $\hpiNN$.
\begin{figure}[tbp]
\begin{center}
\epsfig{file=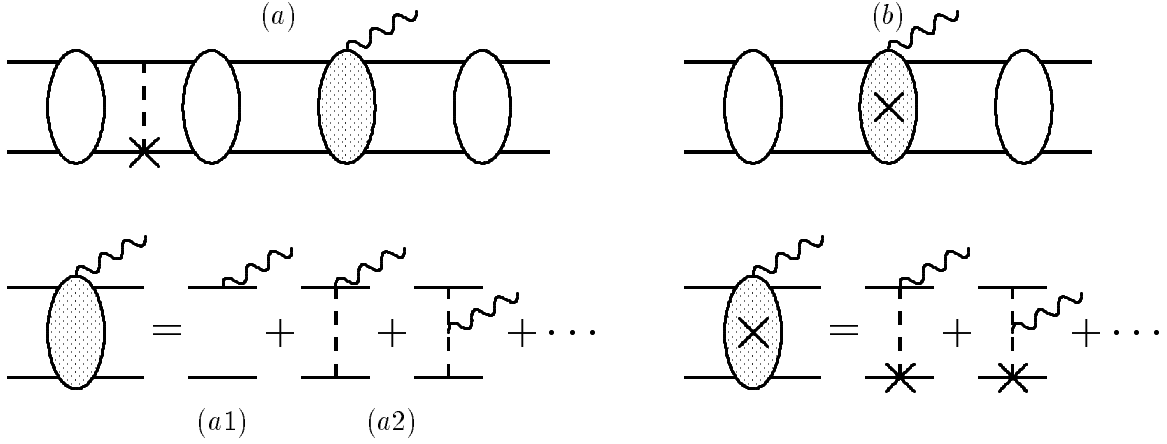} 
\caption[]{Generic diagrams that contribute to the asymmetry.
The solid, dashed and curly lines are nucleons, pions and photons;
and the ``$\times$'' marks denote the insertion of the PNC vertex.
Empty blobs stand for the sum of arbitrary number of iterations
of the parity-conserving strong interactions,
while the shaded blobs with (without) the ``$\times$'' mark do
the PNC (PC) two-nucleon irreducible currents.
None of the crossing diagrams have been drawn.
The shaded blob in $(a)$ will be denoted as
${\vec J}$,
while that in $(b)$ will be denoted by $\delta {\vec J}$.
\label{fig:generic}
}
\end{center}
\end{figure}
Generic diagrams for the asymmetry are
drawn in Fig.~\ref{fig:generic},
where the first and the last terms in eq.(\ref{dpsi})
correspond to Fig.~$(a)$ and Fig.~$(b)$, respectively.
The second term, being symmetric to the first, is not shown
in the figure. Note that the final state wavefunction
in Fig.~$(a)$ has a parity-violating vertex.

As mentioned, we follow Weinberg's power counting \cite{wein90},
where an irreducible diagram is counted as
of order $(Q/\Lambda_\chi)^\nu$,
$Q$ is the typical momentum scale and/or pion mass
and $\Lambda_\chi \sim 4\pi f_\pi \sim m_N$ is the
chiral scale.
The chiral index $\nu$ is given as
$$\nu= 2 L - 2( C -1) 
-1 + \sum_i \nu_i,$$
where $C$ and  $L$ are
the numbers of the separate pieces and loops, respectively.
A vertex indexed by the subscript $i$ is characterized by
$\nu_i \equiv d_i + \frac{n_i}{2} + e_i - 2$,
where $d_i$, $n_i$ and $e_i$ are the number of
derivatives/$m_\pi$'s, nucleon lines and the
external fields, respectively.
For PC interactions,
chiral symmetry guarantees $\nu_i \ge 0$ \cite{rho-prl91}.
Fig.~$(a1)$ is LO with $C=2$,
while Fig.~$(a2,b)$ are NLO with $C=1$.
However, the counting rules for the process at hand is tricky.
For example, the PNC amplitudes due to
Fig.~$(a)$  contain the two-nucleon reducible part,
while the above counting rule is for irreducible diagrams.
The unnatural smallness of the binding energy of the deuteron,
$B_d$, can also contaminate the counting rule,
in the sense that a quantity proportional to positive powers
of $B_d$ becomes much smaller than what the counting rule implies.
We find that the convergence of the chiral expansion is rather poor
and we should go to at least up to NLO to have accurate results.

To be specific in convention, we also write down
the parity-conserving (PC) strong lagrangian
explicitly,
\be
{\cal L}_{pc} = - \frac{g_A}{2 f_\pi} {\bar N} 
 {\vec \sigma}\cdot{\vec \nabla} \pi^i \tau^i N
+ \cdots,
\label{eq:pcpinn}\ee
with $g_A \simeq + 1.26$, $f_\pi\simeq +93\ \mbox{MeV}$
and the ellipsis denotes terms not relevant in the discussion.
Eqs.(\ref{eq:pncpinn}, \ref{eq:pcpinn}) then lead to
the PNC one-pion-exchange (OPE) potential 
that is identical to the one given in \cite{des75, dmnpa78, ddh80},
\begin{eqnarray}
V_{pnc}(\vec r)= \frac{\hpiNN g_A}{2\sqrt{2} f_\pi}
\left(\vec{\tau}_1 \times \vec{\tau}_2\right)^z
\left(\vec{\sigma}_1 + \vec{\sigma}_2\right) \cdot \hat{r} \,
\frac{d}{dr}\left(\frac{e^{-m_\pi r}}{4 \pi r}\right),
\label{VPNC}
\end{eqnarray}
where $\vec{r}\equiv \vec{r}_1 - \vec{r}_2$,
$r\equiv |\vec{r}|$ and $\hat{r}\equiv \vec{r}/r$.
Note that when Siegert's theorem is used,
the sign of the 
asymmetry is determined by the PNC potential.
Thus the consistency of our convention to the
conventional studies are guaranteed by the above PNC
potential.
\footnote{KSSW \cite{kssw99} used PC Lagrangian
with different sign, which causes different sign in the
PNC potential and the asymmetry.}

The one-body currents (Fig.~$(a1)$) read
\be
{\vec J}_{\rm 1B} = 
 \sum_{i=1}^2 \left[
 \frac{1+\tau_i^z}{4} \frac{{\vec p}_i}{m_N} 
 - i \frac{\mu_S + \mu_V \tau^z_i}{4 m_N} 
      \vec{\sigma}_i \times {\vec k}_\gamma
 + \cdots \right],
\ee
where $\mu_S = \mu_p+\mu_n \simeq 0.880$,
$\mu_V = \mu_p-\mu_n\simeq 4.706$
and the ellipse denotes higher order terms.
The two-body PC (Fig.~$(a2)$) and PNC (Fig.~$(b)$) 
currents read
\bea
{\vec J}_{\rm 2B}
&=& \int \frac{d^3{\vec q}}{(2\pi)^3} \e^{i \vq \cdot \vr}\,
 i (\vtau_1\times\vtau_2)^z \frac{g_A^2}{4 f_\pi^2}
 \frac{\del}{\del \vq} \frac{\vsigma_1\cdot \vq\,\vsigma_2\cdot\vq}{
 m_\pi^2+\vq^2}
\nonumber \\
&=& - \vec{r}
 \frac{g_A^2}{12 f_\pi^2}
(\vtau_1\times\vtau_2)^z
 \left[
 \vsigma_1\cdot\vsigma_2 \left( m_\pi^2 y_0(r) - \delta^{(3)}(\vr)\right)
 + S_{12}(\hat{r}) y_2(r) \right],
\\
\delta \vec{J}_{\rm 2B}
&=&
- \int \frac{d^3{\vec q}}{(2\pi)^3} 
{\rm e}^{i\vec{q}\cdot\vec{r}}
\left[\vec{\tau}_1\cdot\vec{\tau}_2-\tau_1^z\tau_2^z\right]
\frac{\hpiNN}{2\sqrt{2}}\frac{g_A}{f_\pi}
\frac{\partial}{\partial\vec{q}}
\frac{(\vec{\sigma}_1  + \vec{\sigma}_2)\cdot \vec{q}}{m^2_\pi + \vec{q}^2}
\nonumber \\
&=& - \hat{r}\ \frac{\hpiNN}{2\sqrt{2}}\frac{g_A}{f_\pi}
\left[\vec{\tau}_1\cdot\vec{\tau}_2-\tau_1^z\tau_2^z\right]
(1+m_\pi r) y_0(r)
\hat{r}\cdot (\vec{\sigma}_1+\vec{\sigma}_2)
\label{JPNC}
\eea
where
\be
y_2(r) \equiv r \frac{\del}{\del r} \frac{1}{r} \frac{\del}{\del r}
y_0(r),
\ \ \
y_0(r)\equiv \frac{\e^{-m_\pi r}}{4\pi r}
\ee
and
\[ S_{12}(\hat{r}) \equiv 
3 \vec{\sigma}_1 \cdot \hat{r}
\vec{\sigma}_2 \cdot \hat{r}
-  \vec{\sigma}_1 \cdot \vec{\sigma}_2.
\]
The PC currents,
${\vec J}_{\rm 1B}$ and ${\vec J}_{\rm 2B}$,
contribute to the asymmetry through the small parity-odd components
of the wavefunctions
induced by the PNC potential, as drawn Fig.~$(a)$.
Here we recall that the PNC potential eq.(\ref{VPNC})
does not commute with either isospin
or orbital angular momentum
and induces parity-odd components 
(which are linearly proportional to $\hpiNN$)
in the spin-triplet wavefunctions. 
On the other hand, the PNC currents, $\delta {\vec J}_{\rm 2B}$,
contribute to the asymmetry by
connecting the parity-even components.

At threshold, there are only $S$-waves in the initial $np$ states,
$\Psi^{000}_{0}$ (${}^1S_0$) and
$\Psi^{011}_{J_z}$ (${}^3S_1$),
where $\Psi^{LSJ}_{J_z}$ are the partial waves.
We write the wavefunctions as
\bea
\Psi^{000}_0(\vec{r})&=& \frac{1}{\sqrt{4\pi} r} u_s(r) \zeta_{10} \chi_{00},
\nonumber \\
\Psi^{011}_{J_z}(\vec{r}) &=& \frac{1}{\sqrt{4\pi} r}
\left(u_t(r) + 
\frac{S_{12}(\hat{r})}{\sqrt{8}} w_t(r) \right)
\zeta_{00} \chi_{1 J_z},
\nonumber \\
\delta \Psi^{011}_{J_z}(\vec{r}) &=& 
-i \sqrt{\frac38}\,
\frac{1}{\sqrt{4\pi} r} 
\left(\vec{\sigma}_1 + \vec{\sigma}_2\right) \cdot \hat{r}
 v_t(r) \zeta_{10} \chi_{1 J_z},
\label{eq:fullwave}
\eea
where $\chi$($\zeta$) represents spinor(isospinor).
We multiplied ``$i$" in front of the $P-$wavefunction
to make $v_t(r)$ real at threshold.
The deuteron wavefunction is the same with $\Psi^{011}$
but with the subscript  ``$d$'', instead of ``$t$''.
The radial functions are normalized as
$\lim_{r\rightarrow \infty} u_s(r) = r - a_s$,
$\lim_{r\rightarrow \infty} u_t(r) = r - a_t$
and 
$\lim_{r\rightarrow \infty} u_d(r) = \mbox{e}^{-\gamma_d r}$,
where
$a_s\ \simeq -23.7\ \mbox{fm}$ 
($a_t\simeq 5.42\ \mbox{fm}$) 
is the spin-singlet (spin-triplet) $np$ scattering length
and $\gamma_d=\sqrt{B_d m_N}$.
The boundary conditions for the ${}^3P_1$ radial wavefunctions
are
$\lim_{r\rightarrow \infty} v_t(r) \propto \hpiNN\,\frac{1}{r^2}$
and 
$\lim_{r\rightarrow \infty} v_d(r) \propto \hpiNN\, 
 \left(1 + \frac{1}{\gamma_d r}\right) \e^{-\gamma_d r}$.

It is well-known (see, for example, \cite{pmr-prl95}) that
the total cross section near threshold
is predominated
by the isovector M1 transition,
${}^1S_0\rightarrow d$,
\be 
\langle \Psi_d | {\vec J} | \Psi_0^{000}\rangle
= \chi^\dagger_{1 M_d}\left[-i 
   (\vec{\sigma}_p - \vec{\sigma}_n) 
\times \hat{k}\, \mbox{M}({}^1S_0)\right] \chi_{00}
\ee
with
\be
\mbox{M}({}^1S_0) = 
 \frac{\omega\mu_V}{4 m_N} 
 \int_0^\infty\! dr \,u_d(r)\ u_s(r) + \cdots
 = (1 + \delta_{\rm 2B})\ (0.263 \ {\mbox{fm}}^2)
\ee
where ${\vec k}=\omega \hat{k}$ is the momentum carried out by the photon,
the ellipsis denotes the two-body-current contributions and
$\delta_{\rm 2B} = (4.6 \pm 0.3)\ \%$ denotes
the ratio of the two-body currents compared to the one-body
contribution.
For the transition from the spin-triplet $np$ state,
there are both PC isoscalar M1 and PNC E1 contributions,
\begin{eqnarray}
\langle \Psi_d | {\vec J}
| \Psi_{J_z}^{011}\rangle 
&=& \chi^\dagger_{1 M_d}\left[-i 
(\vec{\sigma}_p + \vec{\sigma}_n) \times\hat{k}\, \mbox{M}({}^3S_1)
   \right] \chi_{1 J_z},
\nonumber \\
\delta \langle \Psi_d | {\vec J}
| \Psi_{J_z}^{011}\rangle 
&=& \chi^\dagger_{1 M_d}\left[
(\vec{\sigma}_p + \vec{\sigma}_n) \mbox{E}({}^3S_1)
 \right] \chi_{1 J_z}.
\end{eqnarray}
The isoscalar M1 transition is 
tiny (less than $0.1\ \%$) in the total cross section
and does not contribute to the asymmetry.
Detailed analysis can be found in \cite{pkmr2000, crs99}
and we will neglect this isoscalar contribution
hereafter.

The asymmetry reads
\be
A_\gamma = -2 \frac{\mbox{E}({}^3S_1)}{\mbox{M}({}^1S_0)},
\ee
where the $\mbox{E}({}^3S_1)$ is
the consequence of the PNC interaction
and proportional to $\hpiNN$.
The LO $\mbox{E}(^3S_1)$ comes from
the one-body (1B) contribution (Fig.~$(a1)$),
while the two-body contributions
(Fig.~$(a2)$) and $(b)$) are NLO,
\be
\mbox{E}(^3S_1)
=\hpiNN\ \barE,
\ \ \
\barE=\barE_{\rm 1B} +\barE_{\rm 2B}
\ee
with
\begin{eqnarray}
\barE_{\rm 1B}
 &=& - \frac{1}{2\sqrt{6} m_N}
\int_0^\infty\! dr \ \left[
\tilde{v}_d^\prime(r) \left( u_t(r) + \frac{w_t(r)}{\sqrt{2}}\right)  
+ \frac{\tilde{v}_d(r)}{r} \left( u_{t}(r) - \sqrt{2} w_{t}(r) \right)
+ (d\leftrightarrow t)
\right],
\nonumber \\
\barE_{\rm 2B}
&=&
- \frac{g_A^2}{12\sqrt{6} f_\pi^2}
\int_0^\infty\! dr \, r \left[ 
m_\pi^2 y_0(r) - \delta^{(3)}(\vr) + 2 y_2(r) \right]
\left[ \left(u_d(r) + \frac{w_d(r)}{\sqrt{2}}\right) \tilde{v}_t(r)
+ (d\leftrightarrow t)
\right]
\nonumber \\
&+&
\frac{g_A}{3 \sqrt{2} f_\pi}
\int_0^\infty\! dr y_0(r) (1+m_\pi r)
\left(u_d(r) + \frac{w_d(r)}{\sqrt{2}}\right) 
\left(u_t(r) + \frac{w_t(r)}{\sqrt{2}}\right),
\label{barE}
\end{eqnarray}
where $\tilde{v}_{d,t}(r)$ is defined by
$\tilde{v}_{d,t}(r) \equiv v_{d,t}(r)/\hpiNN$
and the ``$(d\leftrightarrow t)$'' denotes the permutation
between
the subscript ``$d$'' and ``$t$''.
With Argonne $v_{18}$ wavefunctions
we have
$\barE= (0.0428 - 0.0302)\ {\mbox{fm}}^2 = 0.0136\ {\mbox{fm}}^2$ 
and consequently
\begin{eqnarray}
A_\gamma^{\rm ChPT}  = 
-0.10\ \hpiNN\,.
\label{eq:result}
\end{eqnarray}

It might be worthwhile making a comparison of
our result eq.(\ref{eq:result}) with
the Siegert's theorem prediction,
\be
\barE^{Siegert}
 = \frac{\omega}{4\sqrt{6}}
\int_0^\infty\! dr\ r \ \left[
  \tilde{v}_d(r) \left( u_{t}(r) + 
\frac{w_{t}(r)}{\sqrt2} \right)
- (d\leftrightarrow t)
\right] = 0.0148\ {\rm fm}^2\,,
\label{FS}\ee
where the numerical value is obtained with
Argonne $v_{18}$ wavefunctions.
Our result is close to the Siegert's theorem prediction,
but the chiral convergence is rather slow,
\be
\frac{ \barE_{\rm 1B}+ \barE_{\rm 2B}}{\barE^{Siegert}}
= 2.89 -1.97 = 0.92\,.
\ee
While a more study is needed to be definite,
here let us present a plausible scenario for the bad convergence.
One can see that
the Siegert's theorem prediction, eq.(\ref{FS}),
is suppressed by the smallness of $\omega\simeq B_d$,
which is smaller than its {\it natural} size,
$Q^2/\Lambda_\chi \sim m^2_\pi/m_N \sim 21$ MeV,
by about 10 times.
The suppression mechanism is, however, not manifest in our
HBChPT results, eq.(\ref{barE}).
To understand the consequences of this, 
let us expand $\barE_{\rm 1B}$ and $\barE_{\rm 2B}$ with respect to
$B_d$ at threshold,
\be
\barE_{n\rm B}= A_{n\rm B}^{(0)} + B_d A_{n\rm B}^{(1)}
 + {\cal O}(B_d^2),
\ \ \
(n=1,\,2).
\ee
Then what Siegert's theorem tells us is that
\be
A_{\rm 1B}^{(0)} + A_{\rm 2B}^{(0)} =0.
\label{AB0}\ee
In case both $A_{\rm 1B}^{(0)}$ and $A_{\rm 2B}^{(0)}$ are not zero,
the above equation indicates that
the ratio $\barE_{\rm 2B}/\barE_{\rm 1B}$
becomes $-1$ at $B_d \rightarrow 0$ limit,
which is not much far from our result,
$\barE_{\rm 2B}/\barE_{\rm 1B} \simeq -2/3$.
In this respect the net result eq.(20) is the result of strong 
cancellation between the $B_d$-independent terms,
$A_{n\rm B}^{(0)}$.
By the same token, the above scenario says that
our PNC E1 amplitudes can be contaminated 
by a small deviation from eq.(\ref{AB0}),
when $B_d$ is quite small.
However it should be understood clearly that
the HBChPT results are to be improved systematically
by taking the higher order.
Thus, for example, to include
the two-body vector charge contributions
which have been neglected in Siegert's theorem,
we should rely on a systematic EFT like HBChPT.
It is quite promising to observe that
HBChPT up to NLO could already
explain the $A_\gamma/\hpiNN$ ratio
within 10\ \% (compared to the Siegert's theorem prediction).
We would like to make it clear that
the situation is completely different in, for example,
M1 transition amplitude,
where the suppression factor $\omega\simeq B_d$ can be factored out
and we are left with the quantity which is non-zero even at
$B_d=0$ limit.
In this case, a beautiful chiral convergence has been observed
\cite{pmr-prl95}.

So far, we have limited ourselves to the threshold
limit, where only $S$ $np$ states are relevant.
At thermal energy, there is tiny but non-zero
contribution from the
$np$ $\Psi^{111}$ state.
This ${}^3P_1$ state
gives non-zero E1 transition amplitude
that is independent of the $\hpiNN$ and
dependent on $\hat{p}$,
the direction of the relative $np$ momentum.
This small $P-$component gives rise to PC asymmetry and 
its magnitude was calculated in \cite{cgp97}.
The PC asymmetry arises from the PC scalar
$\hat{n} \cdot (\hat{p} \times \vec{k}_\gamma)$.
While PNC asymmetry measures up-down asymmetry,
PC asymmetry is the left-right asymmetry.
The amount of the PC asymmetry is reported to be 
$\sim 7 \times 10^{-9}$
which can contaminate the exact measurement of $A_\gamma$.
This corresponds to the accuracy goal of
the LANSCE of the order $10^{-9}$. 
Experimental considerations to discriminate these false
signals were illustrated in \cite{snowetal98}.

\vskip 15pt

We thank B. Desplanques for his valuable discussions
during this work. 
TSP is grateful to S.-i. Ando for his comments.
Work of DPM and CHH is partially supported by the 
KOSEF grant No. 1999-2-111-005-5,
KRF grant 1999-015-DI0023, 2000-015-DP0072 and BK21 program.

\end{document}